\documentclass[11pt]{article}
\usepackage
{
        amssymb,
        amsthm,
        amsmath,
        nicefrac
}
\usepackage[top=1in, bottom=1in, left=1in, right=1in]{geometry}
\usepackage{harvard}
\usepackage{nicefrac}
\title{Local Search is Better than Random Assignment \\for Bounded Occurrence Ordering $k$-CSPs\footnote{The conference version of
this paper appeared at STACS 2013.}}
\author{Konstantin Makarychev\\Microsoft Research, Redmond, WA 98052}
\date{}

\newcommand{\OpenFrame}{\rule{0pt}{12pt} \hrule height 0.8pt \rule{0pt}{1pt} \hrule height 0.4pt \rule{0pt}{6pt}}
\newcommand{\CloseFrame}{\rule{0pt}{1pt}\hrule height 0.4pt \rule{0pt}{1pt} \hrule height 0.8pt \rule{0pt}{12pt}}

\newcommand {\indicator} {{\text{\small{\textbf{1}}}}}
\DeclareMathOperator {\argmax}  {argmax}

\DeclareMathOperator {\val} {value}

\DeclareMathOperator {\sgn}  {sgn}

\newcommand {\Sym}  {\mathfrak{S}}

\newcommand {\Avg}{\text{\sc Avg}}
\newcommand {\Opt}{\text{\sc Opt}}
\newcommand {\Alg}{\text{\sc Alg}}
\newcommand {\Wst}{\text{\sc Worst}}

\newcommand {\roundup}   [1] {{\lceil {#1} \rceil}}

\newcommand {\Exp}       {\mathbb{E}}

\newcommand {\bbR}    {\mathbb{R}}

\newcommand {\calC}   {{\cal{C}}}
\newcommand {\calU}   {{\cal{U}}}

\newtheorem{theorem}{Theorem}[section]

\newtheorem{remark}{Remark}[section]

\begin{document}
\maketitle
\begin{abstract}
We prove that the Bounded Occurrence Ordering $k$-CSP Problem is not approximation
resistant. We give a very simple local search algorithm that always performs better than the random 
assignment algorithm (unless, the number of satisfied constraints does not depend on
the ordering). Specifically, the expected value of the solution returned by the algorithm is at least
$$\Alg \geq \Avg + \alpha (B,k) (\Opt-\Avg),$$
where $\Opt$ is the value of the optimal solution; $\Avg$ is the expected value of the random solution; and $\alpha(B,k)=\Omega_k(B^{-(k+O(1))})$ is
a parameter depending only on $k$ (the arity of the CSP) and  $B$ (the maximum number of times each variable
is used in constraints).

The question whether bounded occurrence ordering $k$-CSPs are approximation resistant was raised by \citeasnoun{GZ12},
who recently showed that bounded occurrence 3-CSPs and ``monotone'' $k$-CSPs admit a non-trivial approximation.
\end{abstract}
\section{Introduction}
{\bf Overview. }
In this work, we give a very simple local search algorithm for ordering constraints satisfaction 
problems that works better than the random assignment for those instances of the ordering $k$-CSP problem, where 
each variable is used only a bounded number of times. To motivate the study of the problem,
we first overview some known results for regular constraint satisfaction problems.

An instance of a constraint satisfaction problem consists of a set of variables $V = \{x_1,\dots, x_n\}$
taking values in a domain $D$ and a set of constraints $\calC$. Each constraint $C\in \calC$ is a function 
from $D^k$ to $\bbR^+$ applied to $k$ variables from $V$. Given
an instance of a CSP, our goal is to assign values to the variables to maximize the 
total payoff of all constraints:
$$\max_{x_1,\dots,x_n\in D^n} \sum_{C\in \calC} C(x_1,\dots, x_n).$$
Note, that we write $C(x_1,\dots, x_n)$ just to simplify the notation. In fact, $C$ may depend on at most $k$ 
variables. The parameter $k$ is called the arity of the CSP. In specific CSP problems, 
constraints $C$ come from a specific family of constraints. For example,
in Max Cut, the domain is $D=\{-1,1\}$, and all constraints have the form $C(x_1,\dots,x_n) = \indicator(x_i\neq x_j)$; in Max 3LIN-2,
the domain $D=\{0, 1\}$, and 
all constraints have the form 
$C(x_1,\dots,x_n) = \indicator(x_i\oplus x_j \oplus x_l = 0)$ 
or $C(x_1,\dots,x_n) = \indicator(x_i\oplus x_j \oplus x_l = 1)$.

Various approximation algorithms have been designed for CSPs. The most basic among them, the ``trivial'' probabilistic algorithm simply assigns random values 
to the variables. It turns out, however, that in some cases this algorithm is essentially optimal. \citeasnoun{Hastad97} showed that for some CSPs e.g., 3LIN-2 and 
E3-SAT, beating the approximation ratio of the random assignment algorithm (by any positive constant $\varepsilon$) is NP-hard. 
Such problems are called approximation resistant. That is, a constraint satisfaction problem is approximation resistant, if for
every positive $\varepsilon > 0$, it is NP-hard to find a $(A_{trivial}+\varepsilon)$ approximation, where $A_{trivial}$ is the approximation ratio of the random
assignment algorithm. If there exists an algorithm with the approximation ratio $(A_{trivial}+\varepsilon)$ for some positive $\varepsilon$, we
say that the problem {\em admits a non-trivial approximation}. It is still not known which constraint satisfaction problems are approximation resistant and which admit a non-trivial 
approximation. This is an active research question in approximation algorithms. 

Suppose now that in our instance of $k$-CSP, each variable is used by at most $B$ constraints. (For example, for Max Cut, this means that the maximum 
degree of the graph is bounded by B.)  \citeasnoun{Hastad00} proved that such instances (which we call $B$-{\em bounded occurrence }$k$-CSPs)
admit a non-trivial approximation. Let $\Opt$ denote the value of the optimal solution; 
$\Avg$ denote the expected value of the random assignment; and $\Alg$ denote the expected value returned by the algorithm. 
\citeasnoun{Hastad00} showed that there exists an approximation algorithm such that\footnote{The quantity $(\text{``value of the solution''} - \Avg)$ is 
called the {\em{advantage over random}}. The algorithm of \citeasnoun{Hastad00} is $O_k(B)$ approximation algorithm for the advantage over random:
$$\Alg - \Avg \geq \frac{\Opt - \Avg}{O_k(B)}.$$}
$$\Alg \geq \Avg + \frac{\Opt - \Avg}{O_k(B)}.$$
Here the hidden constant in $O_k(\cdot)$ may depend on $k$. \citeasnoun{Tre01} showed a
hardness of approximation lower bound of $\Avg + (\Opt - \Avg)/(\Omega_k(\sqrt{B}))$. 

In this work, we study {\em{ordering}} constraints satisfaction problems. 
A classical example of an ordering $k$-CSP is the Maximum Acyclic Subgraph problem. Given
a directed graph $G=(V,E)$, the goal is to find an ordering of
the vertices $\pi: V\to \{1,\dots,n\}$ ($\pi$ is a bijection; $n=|V|$), so as to 
maximize the number of forward edges. In this case, the edges of the
graph are constraints on the ordering $\pi$. An edge $(u,v)$ corresponds
to the constraint $\pi(u)<\pi(v)$. Another example is 
the Betweenness problem. We are given a set of vertices $V$ and a set of constraints $\{C_{u,v,w}\}$.
Each $C_{u,v,w}$ is defined as follows: 
$C_{u,v,w}(\pi) = 1$, if $u<v<w$ or $w<v<u$, and $C_{u,v,w}(\pi) = 0$, otherwise. The goal again 
is to find an ordering satisfying the maximum number of constraints.

More generally, in an ordering $k$-CSP, each constraint $C$ is a function of the ordering
that depends only on the relative order of $k$ vertices. The goal is given a set of vertices $V$ and a 
set of constraints $\calC$, to find an ordering $\pi: V\to [n]$ to 
maximize the total value of all constraints:
$$\max_{\pi:V\to[n]} \sum_{C\in \calC} C(\pi).$$
Here $\pi$ is a bijection and $n=|V|$. If all constraints take values $\{0,1\}$, then the objective is simply to
maximize the number of satisfied constraints. Note, that an {\em ordering} $k$-CSP is not a $k$-CSP.

Surprisingly, we know more about ordering CSPs than about regular CSPs. 
\citeasnoun{GHM11} showed that every ordering CSP problem is approximation resistant assuming the 
Unique Games Conjecture (special cases of this result were obtained by  \citeasnoun{GMR08} and \citeasnoun{CGM09}). On the positive side, \citeasnoun{BS90} showed that
bounded degree Maximum Acyclic Subgraph, and thus every bounded occurrence ordering 2CSP, admits a non-trivial approximation. Their result implies that
$\Alg \geq \Avg + (\Opt - \Avg)/O(\sqrt{B})$. \citeasnoun{CMM07} showed that a slight advantage over the random 
assignment algorithm can be also achieved for instances of Maximum Acyclic Subgraph ($\Alg \geq \Avg + (\Opt - \Avg)/O(\log n)$)
whose maximum degree is not bounded. \citeasnoun{Gutin} showed that the ``advantage over the random 
assignment'' for ordering 3CSPs is {\em fixed--parameter tractable} (we refer the reader to the paper for definitions and more details).
Finally, \citeasnoun{GZ12} proved that all bounded occurrence ordering 3CSPs admit a non-trivial approximation
($\Alg \geq \Avg + (\Opt - \Avg)/O_k(B)$). They also proved that there exists an approximation algorithm for {\em monotone }$k$-CSP
(i.e., ordering CSPs, where all constraints are of the form 
$\pi(u_{i_1}) < \pi(u_{i_2}) < \dots < \pi(u_{i_k})$) with approximation ratio $1/k! + 1/O_k(B)$.

{\textbf{Our results.}} We show that a very simple randomized local search algorithm finds a 
solution of expected value:
\begin{equation}\label{eq:mainclaim}
\Alg \geq \Avg + \frac{\Opt-\Avg}{O_k(B^{k+2})}.
\end{equation}
This algorithm works for every bounded occurrence ordering $k$-CSP. Consequently, all bounded occurrence ordering $k$-CSPs
admit a non-trivial approximation. The running time of the algorithm is $O(n\log n)$.
We do not know whether the dependence on $B$ is optimal. However, the result of \citeasnoun{Tre01}
implies a hardness of approximation upper bound of $\Alg + (\Opt-\Avg)/\Omega_k(\sqrt{B})$
\footnote{Every $k$-CSP can be encoded by an ordering
$2k$-CSP by replacing every boolean variable $x$ with two variables $u_x^{\leftarrow}$ and $u_x^{\rightarrow}$, and letting $x = 1$ if and only if
$\pi(u_x^{\leftarrow}) < \pi(u_x^{\rightarrow})$.}.

\textbf{Techniques.} Our algorithm works as follows: first, it permutes all vertices in a random order. Then, $n$
times, it picks a random vertex and moves it to the optimal position without changing the positions
of other vertices. We give an elementary proof that this algorithm performs better than the random 
assignment. However, the bound we get is exponentially small in $B$. 

Then, we improve this bound. Roughly speaking, instead of the original problem we consider
the ``$D$-ordering'' problem, where the algorithm puts vertices in $D\approx Bk$ buckets (possibly, in a clever way), then it randomly permutes vertices in each of the buckets, and finally outputs vertices in the first bucket, second
bucket, etc. This idea was previously used by \citeasnoun{CMM07}, \citeasnoun{GHM11}, \citeasnoun{Gutin} and \citeasnoun{GZ12}.
The transition to ``$D$-orderings'' allows us to represent the payoff function as a Fourier series
with relatively few terms. We prove that the $L_1$ weight of all coefficients of the payoff function
is at least $\Avg + \Omega_k(\Opt-\Avg)$ (Note, that the optimal value of the ``$D$-ordering'' problem may be 
less than $\Opt$). Then we show that (a)
for each vertex we can find one ``heavy'' Fourier coefficient $\hat{f}_{S}$; and (b) when the original local search 
algorithm moves a vertex it increases the value of the solution in expectation by at least $\Omega_k(\hat{f}_{S}/B)$. This concludes the proof.

\textbf{Correction.} In the preliminary version of the paper that appeared at arXiv, we proved the main result of the paper,
Theorem~\ref{thm:main}. We also gave an alternative, more complicated algorithm in the Appendix. We erroneously
claimed that the performance guarantee of the alternative algorithm is slightly better than~(\ref{eq:mainclaim}). This is not the case. So the best 
bound known to the author is~(\ref{eq:mainclaim}).

\pagebreak

\section{Preliminaries}
An instance of an ordering $k$-CSP problem $(V,\calC)$ consists of a set of vertices $V$ of size $n$, and a set of constraints $\calC$. An ordering of vertices
$\pi: V \to \{1,\dots, n\}$ is a bijection from $V$ to $\{1,\dots, n\}$. Each constraint $C\in \calC$ is a function from the set of all ordering 
$\Sym_V = \{\pi: V \to \{1,\dots, n\}\}$ to $\bbR^+$ that depends on the relative order of at most $k$ vertices. That is, for every $C$ there exists a set $T_C\subset V$ of size at most $k$ such that
if for two orderings $\pi_1$ and $\pi_2$,  $\pi_1(u)< \pi_1(v) \Leftrightarrow \pi_2(u)<\pi_2(v)$
for all $u,v\in T_C$, then $C(\pi_1) = C(\pi_2)$.
The value of an ordering $\pi$ equals
$$\val(\pi, \calC) = \sum_{C\in \calC} C(\pi).$$
We will sometimes write $\val((u_1,\dots u_n), \calC)$ to denote the $\val(\pi, \calC)$ for $\pi: u_i\mapsto i$. We denote the optimal value of the problem by
$\Opt(V,\calC)\equiv \max_{\pi\in \Sym_V} \val(\pi, \calC)$, the average value --- the value returned by the random assignment algorithm ---
by $\Avg(V, \calC) = 1/n! \; \sum_{\pi\in \Sym_V} \val(\pi, \calC)$.

\section{Algorithm}

We now present the algorithm.

\OpenFrame

\textbf{Randomized Local Search Algorithm} 

\medskip

\textbf{Input: }a set of vertices $V$, and a set of constraints $\calC$.

\textbf{Output: }an ordering of vertices $(v_1,\dots,v_n)$.
\begin{enumerate}
\item Randomly permute all vertices.
\item Repeat $n$ times: 
\begin{itemize}
\item Pick a random vertex $u$ in $V$.
\item Remove $u$ from the ordering and insert it at a new location to maximize the payoff. I.e., if $v_1, \dots, v_{n-1}$ is the current 
ordering of all vertices but the vertex $u$, then find a location $i$ that maximizes the
$\val(v_1, \dots, v_{i-1}, u, v_{i+1},\dots v_{n-1}, \calC)$, and put $u$ in  the $i$-th position.
\end{itemize}
\item Return the obtained ordering.
\end{enumerate}
\CloseFrame

\begin{theorem}\label{thm:main}
Given an instance $(V,\calC)$ of a $B$-bounded occurrence ordering $k$-CSP problem, 
the Randomized Local Search Algorithm  returns a solution $\pi_{\Alg}$ of expected value
\begin{equation}\label{eq:to-prove}
\Exp\,\val (\pi_{\Alg}, \calC) \geq  
\Avg(V,\calC) +  \frac{\Opt(V,\calC) - \Avg (V,\calC)}{O_k(B^{k+2})}.
\end{equation}
\end{theorem}
\begin{remark}
In fact, our proof implies a slightly stronger bound: The second term on the right hand side of the inequality~(\ref{eq:to-prove}) can be 
replaced with $(\Opt(V,\calC) - \Wst (V,\calC))/O_k(B^{k+2})$, where $\Wst(V,\calC)$ is the
value of the worst possible solution.
\end{remark}
\begin{proof}
I. We first show using an elementary argument that
$$
\Exp\,\val (\pi_{\Alg}, \calC) \geq  \Avg(V,\calC) +  \alpha(B,k)(\Opt(V,\calC) - \Avg (V,\calC)),
$$
for some function $\alpha(B,k)$ depending only on $B$ and $k$. This immediately implies that every bounded occurrence ordering $k$-CSP 
admits a non-trivial approximation. Then, using a slightly more involved argument we prove the bound~(\ref{eq:to-prove}).

 Observe, that the expected value of the solution after step 1 is exactly equal to $\Avg (V,\calC)$. So we need to estimate
how much local moves at step 2 improve the solution. Let $\Delta_u$ be the maximum possible increase in the value
of an ordering $\pi$, when we move $u$ to another position. In other words, $\Delta_u = \max_{\pi^+,\pi^-} (\val (\pi^+, \calC) - \val (\pi^-, \calC))$, 
where the orderings $\pi^+$ and $\pi^-$ differ only in the position of the vertex $u$. Let $\pi^*$ be the optimal ordering, and $\pi_*$ be the worst possible ordering. 
We can transition from $\pi^*$ to $\pi_*$ by moving every vertex $u$ at most once. Thus,
$$\sum_{u\in V} \Delta_u \geq \val (\pi^*, \calC) - \val (\pi_*, \calC) = 
\Opt(V,\calC) - \Wst(V,\calC)\geq \Opt(V,\calC) - \Avg(V,\calC).$$
Now, our goal is to show that when the algorithm moves a vertex $u$, the value of the solution increases 
in expectation by at least $\alpha(B,k) \Delta_u$ for some function $\alpha$ depending only on $B$ and $k$.

Fix a vertex $u$. Let $\pi^+$ and $\pi^-$ be the orderings that differ only in the position of the vertex $u$ such that
$\Delta_u = \val (\pi^+,\calC) - \val (\pi^-, \calC)$. It may happen that the random permutation chosen by the algorithm 
at step 1 is $\pi^-$, and $u$ is chosen first among all vertices in $V$ at step 2. In this case,
the algorithm can obtain the permutation $\pi^+$ by moving $u$, and thus it can increase the value of the solution
by $\Delta_u$. However, the probability of such event is negligible. It is $1/n\cdot 1/n!$. The main observation is that 
the increase in the value of the ordering, when we move $u$, depends only on the  
order of the {\em neighbors }of $u$ i.e., those vertices that share at least one common constraint $C\in \calC$
with $u$ (including $u$ itself). We denote the set of neighbors by $N(u)$. Since each vertex participates in
at most $B$ constraints, and every constraint depends on at most $k$ variables, $|N(u)|\leq kB$.

Consider an execution of the algorithm. We say that $u$ is {\em fresh }if $u$ was chosen at least once in the ``repeat'' loop of the algorithm, and
none of the neighbors were chosen before $u$ was chosen the first time.  The probability that a variable
$u$ is fresh is at least $\nicefrac{1}{2}|N(u)|^{-1}$. Indeed,
the probability that at least one vertex in $N(u)$ is chosen is  $1 - (1 - |N(u)|/n)^n > 1 - \nicefrac{1}{e}$; the probability that 
the first vertex chosen in $N(u)$ is $u$ is $1/|N(u)|$ (since all vertices in $N(u)$ have the same probability of being chosen first).

If $u$ is fresh, then when it is chosen, its neighbors are located in a random order (since none of them was moved by the algorithm). Thus,
with probability $1/N(u)!\geq 1/(kB)!$, the order of neighbors of $u$ is the same as in $\pi^-$. Then, by moving $u$ we can increase the value of the 
ordering by $\Delta_u$. 

Therefore, when the algorithm moves the vertex $u$, the value of the ordering increases in expectation by at least
$$\Pr(u \text{ is fresh})\cdot \Pr (N(u) \text{ is ordered as }\pi^- 
\text{ after step 1})\cdot \Delta_u = \frac{\Delta_u}{2|N(u)|\,|N(u)|!}\geq  \frac{\Delta_u}{kB\,(kB)!}.$$
This finishes the elementary proof that a positive $\alpha (B,k)$ exists.

\medskip

II. We now improve the lower bound on $\alpha (B,k)$. 
We show that for a fresh variable $u$, the value of the ordering increases in expectation by at least $\Omega_k(B^{-(k+1)})\Delta_u$,
and thus (\ref{eq:to-prove}) holds.
Let $L=\roundup{\log_2 (|N(u)|+1)}$ and $D=2^L$. Consider $D$ buckets $[D] =\{1,\dots, D\}$. For every mapping $x$ of
the vertices to the buckets $v\mapsto x_v \in [D]$, we define a distribution $\calU_x$ on orderings of $V$. A random ordering 
from $\calU_x$ is generated as follows: put each vertex $v$ in the bucket $x_v$; then randomly and uniformly permute vertices in each bucket;
and finally output vertices in the first bucket, second bucket, etc (according to their order in those buckets). Let $\calC_u$ be the set of
constraints that depend on the vertex $u$. Since every variable participates in at most $B$ constraints, $|\calC_u|\leq B$.
Let $f(x)$ be the expected total value of constraints in  $\calC_u$ on a random ordering $\pi$ sampled from 
the distribution~$\calU_x$: 
$$f(x) = \Exp_{\pi\sim \calU_x}\Big[\sum_{C\in\calC_u} C(\pi)\Big].$$

Since the number of buckets $D$ is greater than or equals to $|N(u)|+1$, we may put every vertex in $N(u)$ in its own bucket
and keep one bucket empty. Let 
$\pi^+$ and $\pi^-$ be the orderings as in part I of the proof: $\pi^+$ and $\pi^-$ differ only in the position of the vertex $u$,
and $\val (\pi^+,\calC) - \val (\pi^-, \calC) = \Delta_u$.
 Consider mappings $x^+: V \to [D]$ and
$x^-: V \to [D]$ that put only one vertex from $N(u)$ in every bucket and such that $x_v^+=x^-_v$ for every $v\neq u$,
and $x^+$ orders vertices in $N(u)$ according to $\pi^+$, $x^-$ orders vertices in $N(u)$ 
according to $\pi^-$. For example, if $\pi^+$ arranges vertices in the order $(a,b, u, c)$, 
and $\pi^-$ arranges vertices in the order $(a,u, b, c)$, then 
$x^+ = (a\mapsto 1, *, b\mapsto 3, u\mapsto 4, c\mapsto 5)$ and 
$x^- = (a\mapsto 1, u\mapsto 2, b\mapsto 3, * , c\mapsto 5)$. Since the order of all vertices in $N(u)$ is fixed by $x^+$ and $x^-$,
we have $f(x^+) = \val (\pi^+,\calC_u)$ and $f(x^-) = \val (\pi^-, \calC_u)$. Then
\begin{eqnarray*}
f(x^+) - f(x^-) &=& \val (\pi^+,\calC_u) - \val (\pi^-, \calC_u) \\
&=& \val (\pi^+,\calC) - \val (\pi^-, \calC) = \Delta_u.
\end{eqnarray*}
We now use Theorem~\ref{thm:bound-var}, which we prove in Section~\ref{sec:bound-var}. 
Let $X_v$ (for $v\in V$) be independent random variables uniformly distributed in $[D]$.
By Theorem~\ref{thm:bound-var},
\begin{eqnarray*}
\Exp[\max_{x_u \in D} f(x_u, \{X_v\}_{v\neq u}) - f(X_u, \{X_v\}_{v\neq u})] 
&\geq& \Omega_k(B^{-1}D^{-k})(f(x^+) - f(x^-))\\
&=& \Omega_k(B^{-(k+1)})\Delta_u.
\end{eqnarray*}
Here, $(x_u, \{X_v\}_{v\neq u})$ denotes the mapping $u\mapsto x_u$ and $v\mapsto X_v$ for $v\neq u$;
and $(X_u, \{X_v\}_{v\neq u})$ denotes the mapping $v\mapsto X_v$ for all $v$.

Observe, that when we sample random variables $X_v$, and then sample 
$\pi$ according to $\calU_{X}$, we get a random uniform ordering $\pi$ of all vertices in $V$. Thus,
$$\Exp[f(X_u,\{X_v\}_{v\neq u})] = \Exp_{\pi \in \Sym_V}\Big[\sum_{C\in\calC_u} C(\pi)\Big] = \Exp_{\pi}[\val(\pi, \calC_u)].$$
Similarly, when we sample random variables $X_v$, set $x_u = \argmax_{x_u\in D}f(x_u, \{X_v\}_{v\neq u})$, and then sample $\pi'$ according to $\calU_{(x_u, \{X_v\}_{v\neq u})}$, we get a random uniform ordering of 
all vertices except for the vertex $u$. Denote by $LS(\pi, u)$ the ordering obtained from the ordering $\pi$ by moving the vertex $u$ to the optimal position.
It is easy to see that if $\pi$ is a random uniform ordering, then $LS(\pi, u)$ has the same distribution as $LS(\pi', u)$, since the new
optimal position of $u$ depends only on the relative order of other vertices $v$, and not on the old position of $u$. Hence,
\begin{eqnarray*}
\Exp[\max_{x_u \in D} f(x_u,\{X_v\}_{v\neq u})] &\equiv&
\Exp_{\pi'}[\val(\pi', \calC_u)] \\
&\leq& 
\Exp_{\pi'}[\val(LS(\pi',u), \calC_u)]\\
&=& \Exp_{\pi}[\val(LS(\pi,u), \calC_u)].
\end{eqnarray*}
Hence,
\begin{eqnarray*}
\Exp_{\pi}[\val(LS(\pi,u), \calC) - \val(\pi,u,\calC)] &=& 
\Exp_{\pi}[\val(LS(\pi,u),  \calC_u) - \val(\pi,u, \calC_u)]\\
&\geq& \Omega_k(B^{-(k+1)})\Delta_u.
\end{eqnarray*}
\vskip -0.3mm
\end{proof}

\section{Theorem~\ref{thm:bound-var}}\label{sec:bound-var}
\begin{theorem}\label{thm:bound-var}
Let $D$ be a set of size $2^L$ (for some $L$). Consider a function $f: D^{n+1}\to \bbR$ that can be represented as a sum of $T$ functions $f_t: D^{n+1}\to \bbR$:
$$f(x_0,x_1,\dots, x_n) = \sum_{t=1}^T f_t(x_0, x_1,\dots, x_n)$$
such that each function $f_t$ depends on at most $k$ variables $x_u$. Here, $x_0,\dots, x_n\in D$. Then, the following inequality holds 
for random variables $X_0,\dots, X_n$ uniformly and independently distributed in $D$:
\begin{multline*}
\Exp[\max_{x \in D} f(x,X_1,\dots, X_n) - f(X_0,X_1,\dots, X_n)] \geq \\ 
\Omega_k(T^{-1}|D|^{-k})\max_{x^{+}, x^{-},x_1,\dots, x_n \in D} (f(x^+,x_1,\dots, x_n) - f(x^-,x_1,\dots, x_n)).
\end{multline*}
\end{theorem}
\begin{remark}
The variable $x_0$ corresponds to $x_u$ from the proof of Theorem~\ref{thm:main}. The functions $f_t(x)$ are equal to $\Exp_{\pi\sim \calU_x} C_t(\pi)$,
where $C_t$ is the $t$-th constraint from $\calC_u$.
\end{remark}
\begin{proof}
Without loss of generality we assume that elements of $D$ are vertices of the boolean cube $\{-1,1\}^L$. We denote the $i$-th coordinate 
of $x\in D$ by $x(i)$. We now treat $f$ as a function of $(n+1)L$ boolean variables $x_u(i)$. We write the Fourier series of
the function $f$. The Fourier basis consists of functions
$$\chi_S(x_0,\dots, x_n) = \prod_{(u,i)\in S} x_u(i),$$
which are called {\em characters}. Each index $S\subset \{0,1,\dots,n\}\times \{1,\dots,L\}$ corresponds to the
set of boolean variables $\{x_u(i) :(u,i)\in S\}$. Note, that
$\chi_{\varnothing}(x_0,\dots, x_n) = 1$. The Fourier coefficients of $f$ equal
$$\hat{f}_S = \Exp [f(X_0,\dots, X_n)\, \chi_S(X_0,\dots, X_n)],$$
and the function $f$ equals
$$f(x_0,\dots, x_n) =  \sum_{S} \hat{f}_S\, \chi_S(x_0,\dots, x_n).$$
\begin{remark}
In the proof, we only use the very basic facts about the Fourier transform. The main property we need 
is that the characters form an orthonormal basis, that is, 
$$\Exp [\chi_{S_1}(X_0,\dots, X_n)\chi_{S_2}(X_0,\dots, X_n)] = 
\begin{cases}
1,&\text{if } S_1 = S_2;\\
0,&\text{if } S_1\neq S_2.\\
\end{cases}$$
Particularly, for $S\neq\varnothing$,
$$\Exp [\chi_{S}(X_0,\dots, X_n)] = \Exp [\chi_{S}(X_0,\dots, X_n)\chi_{\varnothing}(X_0,\dots,X_n)] = 0.$$
We will also need the following property: if $f$ does not depend on the variable $x_u(i)$, then 
all Fourier coefficients $\hat{f}_S$ with $(u,i)\in S$ are equal to 0.
\end{remark}

Here is a brief overview of the proof: We will show that the $L_1$ weight of Fourier coefficients of $f$ is at least $f(x^+,x_1,\dots, x_n) - f(x^-,x_1,\dots, x_n)$,
and the weight of one of the coefficients $\hat{f}_{S^*}$ is at least 
$\Omega(T^{-1}|D|^{-k}(f(x^+,x_1,\dots, x_n) - f(x^-,x_1,\dots, x_n)))$. Consequently, if we 
flip a single bit $X_0(i^*)$ in $X_0$ to make the term $\hat{f}_{S^*} \chi_{S^*} (X'_0, X_1, \dots, X_n)$ positive,
we will increase the expected value of $f$ by $|\hat{f}_{S^*}|$.

Observe, that since each function $f_t$ depends on at most $kL$
boolean variables, it has at most $2^{kL} = |D|^k$ nonzero Fourier coefficients. Thus,
$f$ has at most $T\,|D|^k$ nonzero Fourier coefficients~$\hat{f}_S$.

Pick $x^{+}, x^{-},x^*_1,\dots, x^*_n$ that maximize 
$f(x^+,x^*_1,\dots, x^*_n) - f(x^-,x^*_1,\dots, x^*_n)$.
We have 
$$
f(x^+,x^*_1,\dots, x^*_n) - f(x^-,x^*_1,\dots, x^*_n) =
\sum_{S} \hat{f}_S (\chi_S(x^+,x^*_1,\dots, x^*_n) - \chi_S(x^-,x^*_1,\dots, x^*_n)).
$$
If $S$ does not contain pairs $(0,i)$ corresponding to the bits of the variable $x_0$, then 
the character
$\chi_S(x_0,x_1,\dots,x_n)$ does not depend on $x_0$, and 
$\chi_S (x^+,x^*_1,\dots, x^*_n) - \chi_S(x^-,x^*_1,\dots, x^*_n) = 0$, hence
\begin{multline*}
f(x^+,x^*_1,\dots, x^*_n) - f(x^-,x^*_1,\dots, x^*_n) =\\=
\sum_{S: \exists i \text{ s.t. } (0,i)\in S} \hat{f}_S \cdot (\chi_S(x^+,x^*_1,\dots, x^*_n) - \chi_S(x^-,x^*_1,\dots, x^*_n))
\leq 2 \sum_{S: \exists i \text{ s.t. } (0,i)\in S} |\hat{f}_S|.
\end{multline*}
Pick a character $\hat{f}_{S^*}$ with maximum absolute value and pick one of the elements $(0,i^*)\in S^*$.
Since the number of nonzero characters $\hat{f}_S$ is at most $T\,|D|^k$, 
$$|\hat{f}_{S^*}|\geq \frac{f(x^+,x^*_1,\dots, x^*_n) - f(x^-,x^*_1,\dots, x^*_n)}{2T\,|D|^k}.$$

Let $\sigma = \sgn(\hat{f}_{S^*})$. Define a new random variable $X_0'$,
$$X_0'(i)=
\begin{cases}
X_0(i),&\text{for } i\neq i^*;\\
\sigma\, \chi_{S^*}(X_0,\dots,X_n)\, X_0(i),&\text{for } i = i^*.
\end{cases}
$$
 
Consider a character $\chi_S$. If $(0,i^*)\notin S$,
then $\chi_S$ does not depend on the bit $x_0(i^*)$, hence
$\Exp[\chi_S(X_0',X_1,\dots, X_n)] = \Exp[\chi_S(X_0,X_1,\dots, X_n)]$.
On the other hand, if $(0,i^*)\in S$, then 
\begin{multline*}
\Exp[\chi_S(X_0',X_1,\dots, X_n)]= \Exp\big[\frac{X'_0(i^*)}{X_0(i^*)}\,\chi_S(X_0,X_1,\dots, X_n)\big] =\\ 
\Exp[\sigma \chi_{S^*}(X_0,X_1,\dots, X_n) \chi_S(X_0,X_1,\dots, X_n)]= 
\begin{cases}
0,&\text{if } S\neq S^*;\\
\sigma,&\text{if } S = S^*.
\end{cases}
\end{multline*}
The last equality holds because characters $\chi_S$ form an orthonormal basis. Therefore,
$$\Exp[f(X'_0, X_1, \dots, X_n) - f(X_0, X_1, \dots, X_n)] = \sigma \hat{f}_{S^*} = |\hat{f}_{S^*}|.$$
We get
\begin{multline*}
\hskip -0.8mm\Exp[\max_{x \in D} f(x,X_1,\dots, X_n) - f(X_0,X_1,\dots, X_n)] \geq \Exp[f(X'_0, X_1, \dots, X_n) - f(X_0, X_1, \dots, X_n)]
 \\ = 
|\hat{f}_{S^*}| \geq \Omega_k(T^{-1}|D|^{-k})\max_{x^{*}, x_{*},x_1,\dots, x_n \in D} (f(x^+,x_1,\dots, x_n) - f(x^-,x_1,\dots, x_n)).
\end{multline*}

\end{proof}

\pagebreak

\section{Concluding remarks}
We can  guarantee that the algorithm finds a solution of value (\ref{eq:to-prove})
with high probability by repeating the algorithm $\Theta_k(B^{k+2})$ times (since the maxim possible value of the solution is $\Opt$). 

We note that our local search algorithm works not only for ordering $k$-CSPs, but also for (regular) $k$-CSPs. The algorithm first assigns random values
to all variable $x_i$, and then, $n$ times, picks a random $i \in \{1,\dots, n\}$, and changes the value of the variable $x_i$ to the optimal value for fixed other variables. 
The approximation guarantee of the algorithm is $\Avg(V,\calC) +  (\Opt(V,\calC) - \Avg (V,\calC))/O_{k,D}(B)$, here $k$ is the arity, and $D$ is the domain size
of the CSP. The approximation guarantee has the same dependence on $B$ as the approximation guarantee of H{\aa}stad's (2000) original algorithm. 
The analysis relies on Theorem~\ref{thm:bound-var}.

\end{document}